\documentclass[USenglish]{article}
\usepackage[utf8]{inputenc}
\usepackage{authblk}
\usepackage{amsmath,comment}
\usepackage{graphics}
\usepackage{graphicx,color}
\usepackage{amsbsy,amssymb,amsmath,ulem}
\usepackage[all]{xy}

\usepackage{hyperref}

\newcommand{\JJ}{\mathbf{J}}
\newcommand{\XX}{{\mathbf{X}}}
\newcommand{\xx}{{\mathbf{x}}}
\newcommand{\rr}{{\mathbf{r}}}
\newcommand{\uu}{{\mathbf{u}}}
\newcommand{\pd}[2]{\ensuremath{\frac{\partial {#1}}{\partial {#2}}}}
\newcommand{\dd}[2]{\ensuremath{\frac{\mathrm{d} {#1}}{\mathrm{d} {#2}}}}
\newcommand{\MM}{\mathcal{M}}
\newcommand{\bydefinition}{\mathrm{def}}
\newcommand{\absnorm}[1]{\ensuremath{\left|#1\right|}}
\DeclareMathOperator{\divergence}{div}

\begin{document}

\title{Gradient dynamics and entropy production maximization}

\author[1]{Adam Jane\v{c}ka}
\author[1,2]{Michal Pavelka}
\affil[1]{\protect\raggedright 
Mathematical Institute, Faculty of Mathematics and Physics, Charles University in Prague, Sokolovská 83, 186 75 Prague, Czech Republic, e-mail: affro@atrey.karlin.mff.cuni.cz, michal.pavelka@email.cz}
\affil[2]{\protect\raggedright 
Corresponding author: pavelka@karlin.mff.cuni.cz}

\maketitle
\begin{abstract}
Gradient dynamics describes irreversible evolution by means of a dissipation potential, which leads to several advantageous features like Maxwell--Onsager relations, distinguishing between thermodynamic forces and fluxes or geometrical interpretation of the dynamics. Entropy production maximization is a powerful tool for predicting constitutive relations in engineering. In this paper, both approaches are compared and their shortcomings and advantages are discussed.
\end{abstract}




\section{Introduction}
Irreversible phenomena are ubiquitous and it is a goal of non-equilibrium thermodynamics to describe evolution equations governing such processes. There are many frameworks of the non-equilibrium thermodynamics \cite{Lebon-Understanding} leading to countless different ways of prescribing the irreversible terms in evolution equations. We shall discuss two of them, namely the gradient dynamics (dissipation potentials) and the entropy production maximization (EPM).

When using the gradient dynamics, the irreversible part of evolution equations is given by thermodynamic fluxes $\JJ$ which are gradients of a dissipation potential $\Xi$ with respect to conjugate variables or thermodynamic forces, $\XX$:
\begin{equation}\label{eq.JJ}
 \JJ = \frac{\partial \Xi}{\partial \XX}.
\end{equation}
Such framework is advocated, for example, by Edelen \cite{Edelen-AdvChemPhys}, where the non-potential part can be regarded as the reversible part as in \cite{Eringen-CP4}, p. 28, or by recent findings based on the theory of large fluctuations \cite{Mielke-Peletier}. Gradient dynamics also plays a crucial role in the GENERIC framework \cite{GO,OG}, where it guarantees approach to the equilibrium driven by the irreversible part of the evolution equations.

Maximization of entropy production is widely used in both theoretical and engineering communities \cite{Rajagopal2004, ziegler.h:some, ziegler.h.wehrli.c:derivation, ziegler.h.wehrli.c:on}. Constitutive relations (thermodynamic fluxes expressed in terms of forces) are obtained by maximizing a prescribed entropy production $\xi$ subject to constraints such as that the entropy production is a sum of products of thermodynamic forces and fluxes, i.e.,
\begin{equation}\label{eq.EPM}
 \frac{\partial}{\partial \XX}\left( \xi(\XX) + \lambda (\xi(\XX)-\JJ\cdot\XX)\right) = 0,
\end{equation}
$\lambda$ being a Lagrange multiplier. EPM describes many nonlinear irreversible phenomena in particular in non-Newtonian fluids. 

A framework closely related to EPM was developed by G. P. Beretta, see \cite{SEA} and references therein, where entropy production (as sum of products of forces and fluxes) is maximized subject to constraints---the Steepest Entropy Ascent (SEA). One of the constraints is a formula for the entropy production, which is interpreted as a metric on the vector space of fluxes. Therefore, SEA means essentially the following:
\begin{equation}\label{eq.SEA}
 \frac{\partial}{\partial \XX}\left(\JJ\cdot\XX - \tilde{\lambda} \xi(\XX)\right) = 0,
\end{equation}
which is equivalent to Eq. \eqref{eq.EPM}. The two methods (SEA and EPM) are thus essentially equivalent, as shown also in \cite{SEA}. Therefore, relations between gradient dynamics and EPM presented in this paper could be regarded as relations between gradient dynamics and SEA. 

The novel insight brought in this paper is: (i) Comparison of EPM and gradient dynamics and identification of formulas for entropy production for which both approaches coincide. (ii) Explicit identification of a step in the procedure of EPM that is usually tacitly performed, but for which substantial physical insight is necessary. 

\section{Gradient dynamics}
\label{sec.gradient}
We refer to the gradient dynamics as to dynamics generated by a dissipation potential $\Xi$ (sufficiently regular function, zero at the origin and convex near the origin). Let $\xx$ denote the set of state variables.\footnote{For example fields of density, momentum density and entropy density within classical hydrodynamics.} Thermodynamic forces $\XX$ are then related to thermodynamic fluxes $\JJ$ through Eq.~\eqref{eq.JJ}. A relation between thermodynamic forces and fluxes is called a constitutive relation.

\subsection{Legendre transformation}
\label{sec.Leg}

Relation \eqref{eq.JJ} can be seen as a consequence of the Legendre transformation
\begin{subequations}
  \label{eq.LT.XJ}
  \begin{equation}
    \pd{}{\XX} \left( -\Xi (\XX) + \JJ \cdot \XX \right) = 0 \Rightarrow \JJ = \left. \pd{\Xi}{\XX} \right|_{\tilde{\XX}(\JJ)},
  \end{equation}
giving the dependence $\XX = \tilde{\XX}(\JJ)$. The dual dissipation potential is 
\begin{equation}
  \Xi^*(\JJ) = -\Xi ( \tilde{\XX}(\JJ) ) + \JJ \cdot \tilde{\XX}(\JJ). 
\end{equation}
\end{subequations}
The backward Legendre transformation from $\Xi^*(\JJ)$ to $\Xi(\XX)$ proceeds as follows:
\begin{subequations}\label{eq.LT.JX}
\begin{equation}\label{eq.XX}
  \pd{}{\JJ} \left( -\Xi^*(\JJ) + \XX \cdot \JJ \right) = 0 \Rightarrow \XX = \left. \pd{\Xi^*}{\JJ} \right|_{\tilde{\JJ}(\XX)},
\end{equation}
and
\begin{equation}\label{eq.Xi.Xic}
 \Xi(\XX) = - \Xi^* ( \tilde{\JJ}(\XX) ) + \XX \cdot \tilde{\JJ}(\XX).
\end{equation}
\end{subequations}

Legendre transformation is the natural way for passing between $\Xi(\XX)$ and $\Xi^*(\JJ)$ with relation \eqref{eq.JJ} because no information is lost during the passage, see e.g. Callen's textbook \cite{Callen}, where the pertinence of the Legendre transformation in equilibrium thermodynamics is explained. Note also that the forward and backward Legendre transformations need the dissipation potential to be non-degenerate, see \cite{Ogul-TulczyjevI, Ogul-TulzcyjevII} for the degenerate case.

\subsection{Maxwell--Onsager reciprocal relations}
\label{sec.MO}

Onsager reciprocal relations are generalized into far-from-equilibrium regime by using a non-quadratic dissipation potential as shown for example in \cite{PRE2014}. Indeed, taking the derivative of both sides of Eq.~\eqref{eq.JJ} with respect to the force $X_2$, we obtain that
\begin{subequations}
\begin{equation}\label{eq.OCRR}
 \left( \frac{\partial J_1}{\partial X_2} \right)_{X_1} =  
 \left( \frac{\partial J_2}{\partial X_1} \right)_{X_2},
\end{equation}
and change of variables then leads to equivalent relations
\begin{eqnarray}
 \left(\frac{\partial X_1}{\partial J_2}\right)_{J_1} & = &
 \left(\frac{\partial X_2}{\partial J_1}\right)_{J_2}, \\
 \left(\frac{\partial J_1}{\partial J_2}\right)_{X_1} & = &
 -\left(\frac{\partial X_2}{\partial X_1}\right)_{J_2}, \\
 \left(\frac{\partial J_2}{\partial J_1}\right)_{X_2} & = &
 -\left(\frac{\partial X_1}{\partial X_2}\right)_{J_1},
\end{eqnarray}
called the Maxwell--Onsager reciprocal relations (MORR), see \cite{RedExt}.
\end{subequations}

MORR can be also seen as conditions necessary for existence of a dissipation potential generating the dynamics. If they turn out not to be fulfilled, no dissipation potential can be constructed for the given set of thermodynamic forces and fluxes (and state variables). However, a more detailed level of description could be chosen (with an extra state variable), where the dissipative dynamics could be already given by a dissipation potential, and the possible failure of Maxwell--Onsager relations on the less detailed level could be interpreted as an effect of hidden dependence on the missing extra state variable.

\subsection{Identification of forces and fluxes}
Dissipation potential generates irreversible evolution of a set of state variables $\xx$, for example
\begin{equation}\label{eq.dx.irr}
 (\dot{\xx})_{irr} = \left. \pd{\Xi}{\xx^*} \right|_{\xx^* = S_\xx},
\end{equation}
where the derivative is interpreted as a functional derivative. Conjugate state variables $\xx^*$ are identified with derivatives of entropy, see \cite{GO,OG,Adv}. Another possibility is to identify the state variables with derivatives of energy while keeping energy conservation by adding a source term (entropy production) to the balance of entropy density as in \cite{DV}. 

The dissipation potential typically depends only on gradients of the conjugate variables, and thermodynamic forces are then just a shorthand for writing down that dependence, 
\begin{equation}
 \XX = \Gamma (\xx^*)
\end{equation}
where $\Gamma$ is an operator, usually $\Gamma = \nabla$. One than has
\begin{equation}\label{eq.x*.J}
 \Xi_{\xx^*} = -\nabla \cdot \Xi_{\XX} = -\nabla\cdot \JJ.
\end{equation}

For example in classical hydrodynamics, momentum density $\uu$ is among the state variables (together with density and entropy density), and conjugate variables can be identified with derivatives of entropy
\begin{equation}
  S = \int \mathrm{d}\rr\, s \left( \rho, e - \frac{\uu^2}{2\rho} \right)
\end{equation}
where $s$ is the local equilibrium entropy density and $e$ is total energy density. In the isothermal case, conjugate momentum is thus $\uu^* = S_\uu = -\frac{1}{T} \uu/\rho$, which is proportional to the velocity. The corresponding thermodynamic force is the gradient of conjugate momentum, i.e., the velocity gradient. That is why the strain rate tensor is to be interpreted as a thermodynamic force while the irreversible Cauchy stress as the corresponding flux.

\subsection{Entropy production}
Entropy production is given by
\begin{equation}\label{eq.xi}
 \xi = \JJ\cdot \XX = \XX\cdot \Xi_{\XX} = \JJ\cdot \Xi^*_\JJ
\end{equation}
within the gradient dynamics. These relations can be seen from equations \eqref{eq.dx.irr} and \eqref{eq.x*.J}, since the total entropy of a system evolves as
\begin{equation}
 \dot{S} = \int \mathrm{d}\rr\, S_\xx \cdot (\dot{\xx})_{irr} = -\int \mathrm{d}\rr\, S_\xx \cdot \left(\nabla\cdot\JJ\right) = \int \mathrm{d}\rr\, \nabla S_\xx \cdot \JJ = \int \mathrm{d}\rr\, \XX\cdot\JJ.
\end{equation}
Entropy production $\xi$ is thus equal to the product of thermodynamic forces and fluxes as usually in the non-equilibrium thermodynamics \cite{dGM}.

\subsection{Geometric motivation}
Let us now motivate the gradient dynamics from the point of view of differential geometry. All the necessary terminology can be found in \cite{Fecko} and \cite{Marsden-Ratiu}. The set of state variables $\xx$ forms an infinite-dimensional manifold $\MM$, $\xx\in \MM$. Right hand sides of the evolution equations of the state variables are then vector fields on the manifold, which belong to the tangent bundle of the manifold $T\MM$. The vector fields can be split into the reversible and irreversible part as shown in \cite{PRE2014}, and the irreversible part is in close relation with the thermodynamic fluxes, typically their divergence. The fluxes can be thus regarded as elements of the tangent bundle. Each point of the tangent bundle can be thus associated with coordinates $(\xx,\JJ)$.

Cotangent bundle, which is locally dual to the tangent bundle, can be constructed. The construction is analogical to the passage from variables $q,\dot q$ to $q, p$ in the classical mechanics. Let us denote elements of the cotangent bundle as $(\xx, \XX)$. If there is a function on $T\MM$ that expresses the irreversible evolution (dissipation potential $\Xi^*$), what is the corresponding dissipation potential on the cotangent bundle? 

To construct such a function, we first need a mapping from the tangent bundle to the cotangent bundle. Moreover, it should map points $(\xx, \JJ)$ to points $(\xx,\XX)$, where the $\xx$ coordinate is the same, the mapping should preserve the fibers.\footnote{Fiber is the space of all fluxes or forces attached to a particular point $\xx\in \MM$.} Once having a function on $T\MM$, a natural fiber-preserving mapping is the fiber derivative
\begin{equation}
F\Xi^*: (\xx,\JJ) \mapsto \left(\xx, \frac{\partial \Xi^*}{\partial \JJ}\right).
\end{equation}
So the new function on the cotangent bundle should be a function of $\xx$ and of the derivatives of the original function $\Xi^*$. In order not to lose any information (see \cite{Callen}) contained in $\Xi^*$, the new function is given by Legendre transform of the original function, which is given by Eq.~\eqref{eq.LT.JX}. Constitutive relation \eqref{eq.XX} can be thus seen as a natural consequence of the duality between tangent and cotangent bundles. A similar argument based on multiscale thermodynamics was given in \cite{MG-CR}.

In summary, Legendre transformation is the natural transformation between functions on tangent and cotangent bundles, and gradient dynamics can be seen as a consequence of this transformation.

\section{Entropy production maximization}
\label{sec.EPM}

Let us now formulate the procedure of entropy production maximization (EPM). Taking entropy production $\xi(\XX)$, which is a function of thermodynamic forces, the entropy production should be maximized while keeping the constraint
\begin{equation}\label{eq.EPM.constraint}
  \xi = \JJ\cdot \XX.
\end{equation}
The maximization is carried out by means of the Lagrange multipliers technique, 
\begin{subequations}
  \begin{eqnarray}
    \label{eq.EPM.X} 
    \pd{}{\XX} \left( \xi(\XX) + \lambda (\xi - \JJ\cdot\XX) \right) &=& 0, \\
    \pd{}{\lambda} \left( \xi(\XX) + \lambda (\xi - \JJ\cdot\XX) \right) &=& 0,
  \end{eqnarray}
\end{subequations}
where $\lambda$ is the Lagrange multiplier. By solving these two equations, we are able to obtain $\JJ$ as a function of $\XX$. Note that the second equation is equivalent to constraint \eqref{eq.EPM.constraint}.

Equation \eqref{eq.EPM} implies (after multiplying Eq.~\eqref{eq.EPM.X} by $\XX$) that
\begin{equation}
 \frac{1+\lambda}{\lambda} \XX \cdot \frac{\partial \xi}{\partial \XX} = \xi,
\end{equation}
which gives $\lambda$ as a function of $\XX$. Eq. \eqref{eq.EPM.X} then leads to
\begin{equation}
  \label{eq.EPM.J}
  \JJ = \frac{\xi}{\XX\cdot\xi_\XX}\xi_\XX,
\end{equation}
which is the general result (constitutive relation) obtained by the method of entropy production maximization.

In the particular case of $k$-homogeneous entropy production, we have that 
\begin{equation}
\XX\cdot\frac{\partial \xi}{\partial \XX} = k \xi,
\end{equation}
and Eq.~\eqref{eq.EPM.J} becomes
\begin{equation}\label{eq.EPM.J.homo}
 \JJ = \frac{1}{k}\frac{\partial \xi}{\partial \XX}.
\end{equation}
For example the quadratic entropy production
\begin{equation}
  \label{eq.xi.quadratic}
  \xi = \sum_{ij}L_{ij}X_i X_j
\end{equation}
is 2-homogeneous, and Eq.~\eqref{eq.EPM.J.homo} yields the standard linear force-flux relations 
\begin{equation}
 J_i = L_{ij} X_j,
\end{equation}
see \cite{dGM}.

\subsection{Non-uniqueness of the choice of fluxes and forces}
\label{sec:non-uniq-choice}

Let us assume system with two fluxes $J_1$ and $J_2$ which are known functions of the forces $X_1$ and $X_2$. The entropy production is then given by
\begin{equation}\label{eq.EPM.JX}
  \xi (X_1, X_2) = J_1 (X_1, X_2) X_1 + J_2 (X_1, X_2) X_2.
\end{equation}
It could be tempting to identify the thermodynamic fluxes with $J_1$ and $J_2$, but the method of maximum entropy production leads to generally different fluxes $J^*_1$ and $J^*_2$, given by \eqref{eq.EPM.J}. It can be shown that the relation between the known fluxes $J_i$ and the fluxes resulting from the maximum entropy production principle $J^*_i$ is
\begin{subequations}
  \begin{align}
    J^*_1 &= J_1 + \frac{\Delta}{X_1}, \\
    J^*_2 &= J_2 - \frac{\Delta}{X_2},
  \end{align}
\end{subequations}
where the discrepancy $\Delta$ is given by
\begin{equation*}
  \Delta = - \frac{X_1 X_2}{\XX \cdot \xi_\XX} \left( J_1 X_1 \pd{J_1}{X_2} - J_2 X_2 \pd{J_2}{X_1} + J_1 X_2 \pd{J_2}{X_2} - J_2 X_1 \pd{J_1}{X_1} \right).
\end{equation*}
The fluxes are unique if the discrepancy is equal to zero, i.e.,
\begin{equation}
  \label{eq:uniqueness-condition}
  J_1 (X_1, X_2) \pd{\xi}{X_2} = J_2 (X_1, X_2) \pd{\xi}{X_1},
\end{equation}
see for example \cite{martyushev.lm.seleznev.vd:maximum}. For quadratic entropy production, condition \eqref{eq:uniqueness-condition} holds if Onsager reciprocal relations are satisfied.

In summary, it is important to determine the thermodynamic fluxes by going through the whole procedure of entropy production maximization, which leads to formula \eqref{eq.EPM.J}. On the other hand, if the fluxes are identified simply from writing down entropy production in the form $\xi = \JJ(\XX) \cdot \XX$, the result can be misleading, since there are usually more ways of casting entropy production in that form.

\subsection{General framework of the entropy production maximization procedure}\label{sec:procedure}

To find the constitutive relations specifying the system by the means of the entropy production maximization, we need to determine how the system stores energy and how it produces entropy---to this end, we need to stipulate two scalar functions. The general framework goes as follows, for more details see~\cite{malek.j.prusa.v:derivation}.
\begin{description}
\item [STEP 1:] Determine the state variables $\xx$ and specify the storage mechanism of the system by virtue of the fundamental thermodynamic relation in terms of one of the thermodynamic potentials (internal energy $e$, Helmholtz free energy $\psi$, Gibbs potential $G$, enthalpy $h$, \dots) as a function of the state variables. 
\item [STEP 2:] From the balance of internal energy $e$ and the fundamental thermodynamic relation derive the local form of the balance of entropy---the Clausius--Duhem inequality
  \begin{equation}
    \label{eq:balance-of-entropy}
    \rho \dd{\eta}{t} + \divergence \mathbf{j}_\eta = \JJ \cdot \XX > 0,
  \end{equation}
  where $\rho$ is the density, the thermodynamic temperature is defined as $\theta =_\bydefinition \pd{e}{\eta}$, $\mathbf{j}_\eta$ is the entropic flux and $\xi =_\bydefinition \JJ \cdot \XX > 0$ is the entropy production, where the dot product can be understood as a summation of different mechanisms of the entropy production. The non-negativity of the entropy production function $\xi$ is a consequence of the second law of thermodynamics.
\item [STEP 3:] Specify the constitutive relation for the entropy production function in terms of the thermodynamic fluxes $\JJ$ or the thermodynamic forces $\XX$ as
  \begin{equation} 
    \xi = \xi (\JJ), \quad \text{ or } \quad \xi = \xi (\XX).
  \end{equation}
\item [STEP 4:] Maximize the entropy production function $\xi$ with respect to the thermodynamic fluxes $\JJ$ or the thermodynamic forces $\XX$. As a constrain of this maximization procedure, the definition of the entropy production $\xi = \JJ \cdot \XX$ arising form \eqref{eq:balance-of-entropy} must hold. 
  \begin{description}
  \item [STEP 4a:] In case that there is no coupling among individual terms in the entropy production function, i.e., it can be written as
    \begin{equation}
      \label{eq:splitted-ent-production}
      \xi (\JJ) = \sum_\alpha \xi_\alpha (J_\alpha), \quad \text{ or } \quad \xi (\XX) = \sum_\alpha \xi_\alpha (X_\alpha),
    \end{equation}
    we can additionally require that the sought forces or fluxes would depend solely on their corresponding counterparts. Then, we can maximize the particular terms $\xi_\alpha$ separately using the Lagrange multipliers $\lambda_\alpha$ as
    \begin{subequations}
      \begin{align}
        \dd{}{J_\alpha} \left( \xi_\alpha (J_\alpha) + \lambda_\alpha \left( \xi_\alpha (J_\alpha) - J_\alpha X_\alpha \right) \right) &= 0, \qquad \forall \alpha, \\
        \dd{}{\lambda_\alpha} \left( \xi_\alpha (J_\alpha) + \lambda_\alpha \left( \xi_\alpha (J_\alpha) - J_\alpha X_\alpha \right) \right) &= 0, \qquad \forall \alpha,
      \end{align}
    \end{subequations}
    or
    \begin{subequations}
      \begin{align}
        \dd{}{X_\alpha} \left( \xi_\alpha (X_\alpha) + \lambda_\alpha \left( \xi_\alpha (X_\alpha) - J_\alpha X_\alpha \right) \right) &= 0, \qquad \forall \alpha, \\
        \dd{}{\lambda_\alpha} \left( \xi_\alpha (X_\alpha) + \lambda_\alpha \left( \xi_\alpha (X_\alpha) - J_\alpha X_\alpha \right) \right) &= 0, \qquad \forall \alpha.
      \end{align}
    \end{subequations}
    From the same procedure as described in Section~\ref{sec.EPM}, it follows for the particular fluxes or forces
    \begin{equation}
      J_\alpha = \frac{\xi_\alpha}{\dd{\xi_\alpha}{X_\alpha} X_\alpha} \dd{\xi_\alpha}{X_\alpha}, \quad \text{ or } \quad X_\alpha = \frac{\xi_\alpha}{\dd{\xi_\alpha}{J_\alpha} J_\alpha} \dd{\xi_\alpha}{J_\alpha},
    \end{equation}
    where there is no summation over $\alpha$ and we truly have $J_\alpha = J_\alpha (X_\alpha)$ or $X_\alpha = X_\alpha (J_\alpha)$. Note that these expressions cannot be further simplified, since $J_\alpha$ and $X_\alpha$ might be vectorial or tensorial quantities.
  \item [STEP 4b:] When the entropy production cannot be written as~\eqref{eq:splitted-ent-production} or without the additional requirement  made in {\bf STEP 4a}, we need to maximize the entropy production as a whole, thus arriving at the relation~\eqref{eq.EPM.J} or its counterpart for the thermodynamic forces. In this case, the resulting fluxes might depend on all the other forces or the other way around.
  \end{description}
\end{description}
Even though, the requirement made in {\bf STEP 4a} is not necessary, it is usually tacitly considered, since it leads to the commonly used constitutive relations. On the other hand, maximization of the entropy production as a whole leads to rather complicated but possibly richer expressions, see the example in Section~\ref{sec:heat-power}.



\subsection{Summary of EPM}
\begin{figure}[ht!]
\begin{center}
\includegraphics{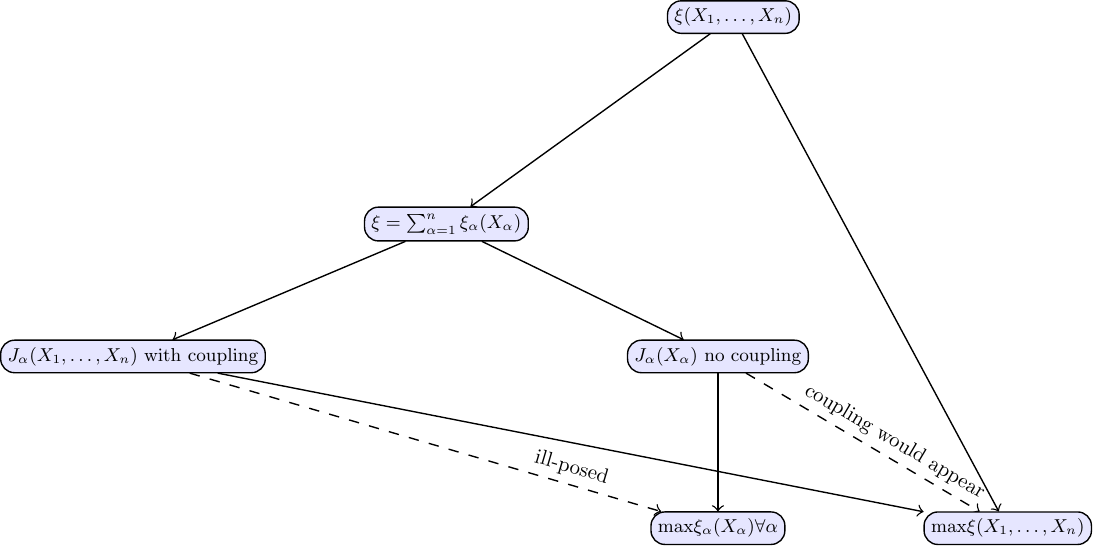}
\caption{Possible routes of EPM. Having a general entropy production $\xi(X_1,\dots,X_n)$, one can directly maximize the whole entropy production. That corresponds to step 4b in Sec. \ref{sec:procedure} of the EPM procedure. If, on the other hand, the entropy production can be split into parts each of which depends only on one force, $\xi = \sum_\alpha \xi_\alpha(X_\alpha)$, which is often the case, one has more options. If coupling between forces is allowed, i.e., the corresponding fluxes depend on all forces in general, the entropy production should be maximized as a whole, otherwise the result would depend on the ambiguous choice of thermodynamic forces, see Sec.~\ref{sec:non-uniq-choice}. If no coupling is allowed, each flux depends only on the corresponding thermodynamic force. One can then infer the constitutive relations by maximizing each part of entropy production separately, which is the most frequent case in the literature. That is the step 4a in Sec.~\ref{sec:procedure}. It should be borne in mind, that even in this case the result depends on the choice of thermodynamic forces as in Sec.~\ref{sec:non-uniq-choice}. However, when coupling is forbidden, the forces can be often separated from each other in a physically plausible way (e.g. separating heat from mechanical stress).
\label{fig.map}}
\end{center}
\end{figure}

The method of entropy production maximization is summarized in Fig.~\ref{fig.map}.

\section{Gradient dynamics vs. EPM}
\label{sec.comparison}

Assume now that the dissipative evolution is described by gradient dynamics as in Sec.~\ref{sec.gradient}, where fluxes can be expressed in terms of forces by means of Eq.~\eqref{eq.JJ}. Entropy production is then given by Eq.~\eqref{eq.xi}, and Eq.~\eqref{eq.EPM.J} becomes
\begin{equation}
  \JJ = \frac{ \XX \cdot \Xi_\XX}{\XX \cdot \left( \Xi_{\XX} + \Xi_{\XX \XX} \XX \right)} (\Xi_{\XX} + \Xi_{\XX \XX} \XX). 
\end{equation}
If the dissipation potential is $k$-homogeneous, see Sec.~\ref{sec.homogeneous}, the last equation becomes Eq.~\eqref{eq.JJ}, and gradient dynamics can be regarded as EPM. 

In the case of only one thermodynamic force $X$ and on the reasonable assumption that the dissipation potential depends on this force through its norm, $\Xi = \Xi \left( \absnorm{X} \right)$, the gradient dynamics also coincides with EPM. Indeed, from~\eqref{eq.xi}, we can compute for the quantities figuring in~\eqref{eq.EPM.J}
\begin{align*}
  \xi &= X \cdot \dd{\Xi}{X} = X \cdot \Xi' \frac{X}{\absnorm{X}} = \Xi' \absnorm{X}, \\
  \dd{\xi}{X} &= \Xi'' X + \frac{\Xi'}{\absnorm{X}} X, \\
  \dd{\xi}{X} \cdot X &= \Xi'' \absnorm{X}^2 + \Xi' \absnorm{X},
\end{align*}
where the prime $(')$ denotes the derivation with respect to the argument---the norm $\absnorm{X}$, and the dot $(\cdot)$ should be understood as a corresponding vector/tensor inner product. Then, we can substitute into~\eqref{eq.EPM.J}
\begin{equation}
  J = \frac{\Xi' \absnorm{X}}{\Xi'' \absnorm{X}^2 + \Xi' \absnorm{X}} \left( \Xi'' + \frac{\Xi'}{\absnorm{X}} \right) X = \frac{\Xi'}{\absnorm{X}} X = \dd{\Xi}{X},
\end{equation}
and we have recovered relation~\eqref{eq.JJ}.

In summary, if the entropy production is either $k$-homogeneous in all the forces or if it depends on only one force, EPM is equivalent to the gradient dynamics.

\subsection{Non-homogeneous entropy production}

\subsubsection{Chemical kinetics}

Consider for example coupled chemical reactions with a dissipation potential
\begin{equation}
  \Xi = k_1 \left[ \cosh \left(\frac{X_1}{2} \right) - 1 \right] + k_2 \left[ \cosh \left(\frac{X_2}{2} \right) - 1 \right],
\end{equation}
where $k_1, k_2$ are constants. Dissipation potential of this form leads to reaction rates
\begin{equation}\label{eq.chemkin.J}
  J_i = \frac{k_i}{2} \sinh \left( \frac{X_i}{2} \right), \qquad i = 1,2.
\end{equation}
This description of chemical kinetics was shown to be compatible with the Guldberg--Waage law of mass action \cite{Grmela2012fluctuations}. From~\eqref{eq.xi}, we see that
\begin{equation}\label{eq.chemkin.xi}
  \xi = \frac{k_1}{2} \sinh \left( \frac{X_1}{2} \right) X_1 + \frac{k_2}{2} \sinh \left( \frac{X_2}{2} \right) X_2,
\end{equation}
and using \eqref{eq.EPM.J} we obtain the reaction rates
\begin{equation}
  \label{eq.chemkin.J.EPM}
  J_i = \frac{k_k \sinh \left( \frac{X_k}{2} \right) X_k}{2 k_l X_l \left[ X_l \cosh \left( \frac{X_l}{2} \right) + 2 \sinh \left( \frac{X_l}{2} \right) \right]} k_i \left[ X_i \cosh \left( \frac{X_i}{2} \right) + 2 \sinh \left( \frac{X_i}{2} \right) \right], \qquad i = 1,2,
\end{equation}
where there is no summation over the index $i$. This does not seem to be the right result due to its complexity and because Eq.~\eqref{eq.chemkin.J} is compatible with the well established law of mass action.

In summary, EPM with the entropy production \eqref{eq.chemkin.xi} leads to thermodynamic fluxes (reaction rates) \eqref{eq.chemkin.J.EPM}, which are different from fluxes \eqref{eq.chemkin.J}. The latter choice of fluxes was however shown to be compatible with the widely accepted law of mass action and Butler--Volmer equation, see \cite{Grmela2012fluctuations} and \cite{Pavelka-AE}, and we thus prefer them to the former choice of fluxes in the case of nonlinear chemical kinetics.

\subsubsection{Incompressible heat-conducting non-Newtonian fluid}
\label{sec:heat-power}

As a simple example, we can consider the following entropy production
\begin{equation}
  \label{eq:heat-power-production}
  \xi = 2 \mu \left( 1 + \alpha \left| \mathbf{D}_\delta \right|^2 \right)^{r-1} \left| \mathbf{D}_\delta \right|^2 + \kappa \left| \nabla \theta \right|^2,
\end{equation}
where $\mu$, $\alpha$ and $\kappa$ are positive constants, $r$ is a constant, $\mathbf{D}_\delta =_\bydefinition \mathbf{D} - \frac{1}{3} \left( \mathrm{Tr}\,\mathbf{D} \right) \mathbf{I}$ is the deviatoric part of the symmetric part of the velocity gradient and $\theta$ is the temperature. The fluxes associated with $\mathbf{D}_\delta$ and the temperature gradient $\nabla \theta$ are the deviatoric part of the Cauchy stress $\mathbf{T}_\delta$ and the negative heat flux $- \mathbf{q}$ respectively, see~\cite{Rajagopal2004} for details. Using the entropy production maximization principle \eqref{eq.EPM.J}, we arrive at
\begin{subequations}
  \label{eq:heat-power-max-fluxes}
  \begin{align}
    \mathbf{T}_\delta &= \frac{2 \mu \left( 1 + \alpha \left| \mathbf{D}_\delta \right|^2 \right)^{r-1} \left| \mathbf{D}_\delta \right|^2 + \kappa \left| \nabla \theta \right|^2}{4 \mu \left( 1 + \alpha \left| \mathbf{D}_\delta \right|^2 \right)^{r-2} \left( 1 + \alpha r \left| \mathbf{D}_\delta \right|^2 \right) \left| \mathbf{D}_\delta \right|^2 + 2 \kappa \left| \nabla \theta \right|^2} 2 (r+1) \mu \left| \mathbf{D}_\delta \right|^{r-1} \mathbf{D}_\delta, \\
    - \mathbf{q} &= \frac{2 \mu \left( 1 + \alpha \left| \mathbf{D}_\delta \right|^2 \right)^{r-1} \left| \mathbf{D}_\delta \right|^2 + \kappa \left| \nabla \theta \right|^2}{4 \mu \left( 1 + \alpha \left| \mathbf{D}_\delta \right|^2 \right)^{r-2} \left( 1 + \alpha r \left| \mathbf{D}_\delta \right|^2 \right) \left| \mathbf{D}_\delta \right|^2 + 2 \kappa \left| \nabla \theta \right|^2} 2 \kappa \nabla \theta, \label{eq:heat-power-max-heat-flux}
  \end{align}
\end{subequations}
which is not particularly neat (and it would be even worse in a hypothetical situation when the viscosity $\mu$ was dependent on the temperature gradient $\nabla \theta$).

Separate maximization of~\eqref{eq:heat-power-production} or using the dissipation potential\footnote{The requirement for the dissipation potential to be convex yields an additional restriction $r \geq \frac{1}{2}$.}
\begin{equation}
  \Xi = \frac{\mu}{\alpha r} \left( 1 + \alpha \left| \mathbf{D}_\delta \right|^2 \right)^r + \frac{1}{2} \kappa \left| \nabla \theta \right|^2,
\end{equation}
leads to a much more luminous relations
\begin{subequations}
  \label{eq:heat-power-diss-fluxes}
  \begin{align}
    \mathbf{T}_\delta &= 2 \mu \left( 1 + \alpha \left| \mathbf{D}_\delta \right|^2 \right)^{r-1} \mathbf{D}_\delta, \\
    - \mathbf{q} &= \kappa \nabla \theta,
  \end{align}
\end{subequations}
where the first equation is the Carreau model, see~\cite{carreau.pj:rheological}, and the second equation is the well-known Fourier law of thermal conductivity.

We see that for $r = 1$, i.e., when both terms in~\eqref{eq:heat-power-production} are quadratic, constitutive relations \eqref{eq:heat-power-max-fluxes} and~\eqref{eq:heat-power-diss-fluxes} are tantamount and we recover the standard model for incompressible heat-conducting Newtonian fluid.

Since there are materials with the thermal conductivity $\kappa$ depending on the shear rate, see for example~\cite{Lee1998}, we hoped that the constitutive relation~\eqref{eq:heat-power-max-heat-flux} given by the entropy production maximization could capture this dependence. Unfortunately, this was not the case neither for the entropy production \eqref{eq:heat-power-production} nor for entropy productions motivated by the Ostwald--de Waele power-law model \cite{ostwald.w:uber,waele.a:viscometry}, the Sisko model \cite{sisko.aw:flow} or the Cross model \cite{cross.mm:rheology}.

In summary, although there is no apparent coupling in entropy production \eqref{eq:heat-power-production}, nonlinear coupling appears after maximization of the whole entropy production. Such a coupling would suggest for example dependence of effective thermal conductivity on shear rate. However, the magnitude of the coupling is not in agreement with experimental observations.

\subsection{Maxwell--Onsager relations}

When discussing compatibility of gradient dynamics and EPM, a question arises whether the Maxwell--Onsager relations from Sec. \ref{sec.MO}, which are necessary for existence of a dissipation potential, are fulfilled when the evolution is obtained by EPM. Taking the derivative of the general equation for flux given by EPM, Eq.~\eqref{eq.EPM.J}, with respect to force $X_2$ and requiring relation \eqref{eq.OCRR} to hold leads to the condition
\begin{equation}\label{eq.EPM.MO}
X_k\left(\frac{\partial^2 \xi}{\partial X_k X_j} \frac{\partial \xi}{\partial X_i}-\frac{\partial^2 \xi}{\partial X_k X_i} \frac{\partial \xi}{\partial X_j}\right) = 0, \qquad \forall i,j.
\end{equation}
If this condition is fulfilled, there may be constructed a dissipation potential describing the evolution and MORR are fulfilled even in the far-from-equilibrium regime. The condition is fulfilled for example for $k-$homogeneous entropy productions. If the condition is not fulfilled, no dissipation potential can be constructed. In particular, the condition is fulfilled for entropy production \eqref{eq.xi.quadratic} when the matrix $L_{ij}$ is symmetric. That means that Onsager reciprocal relations are fulfilled by EPM near equilibrium, where the original Onsager's derivation \cite{Onsager1930,Onsager1931} is formulated. 

Condition \eqref{eq.EPM.MO} is not fulfilled for all entropy productions. For example entropy production
\begin{equation}
 \xi = X_1 \sinh \left( X_1 \right) + X_2 \sinh \left( X^2_2 \right),
\end{equation}
violates it.

Maxwell--Onsager reciprocal relations, which are fulfilled even far from equilibrium by gradient dynamics, are fulfilled by EPM near equilibrium, but not necessarily far from equilibrium. Assuming that a constitutive relation is generated by EPM, MORR can be regarded as the compatibility condition necessary for constructing a dissipation potential leading to the same constitutive relation.


\section{Conclusion}
In section \ref{sec.gradient}, we review gradient dynamics, where irreversible evolution is generated by a dissipation potential. For example, there is a natural way how to distinguish thermodynamic forces and fluxes, missing in the classical non-equilibrium thermodynamic frameworks, e.g. \cite{dGM}. Moreover, Onsager reciprocal relations are automatically extended to far-from-equilibrium regimes and the implied Maxwell--Onsager reciprocal relations are also guaranteed. Gradient dynamics is advantageous because of this automatic consistence with thermodynamics.

In section \ref{sec.EPM}, the method of entropy production maximization (EPM) is recalled and it is compared with gradient dynamics in section \ref{sec.comparison}. Both methods are compatible if entropy production is a homogeneous function of thermodynamic forces or fluxes or if it depends on only one force (or flux). Otherwise the compatibility is rather rare. 

When performing entropy production maximization, a step is usually tacitly made (namely Step 4a in Sec. \ref{sec:procedure}) where entropy production is split into several parts, each of which is maximized separately. In the case of possible coupling between thermodynamic forces, such a step can be done only with great caution, since the resulting coupling is affected by the splitting and since the splitting depends on the not always objective definition of thermodynamic forces, see Sec. \ref{sec:non-uniq-choice}. The step is usually made to separate for example mechanical forces and heat flux, which seems to be natural. One should thus at least mention the non-trivial input into the procedure of EPM when performing the splitting step. 

Instead of splitting entropy production into several independent parts, the whole entropy production can be maximized at once. The resulting constitutive relations then contain nonlinear coupling between the thermodynamic forces. Such coupling, however, seems to be incompatible with experimental observations at least in the case of non-isothermal Couette flow of suspensions, Sec. \ref{sec:heat-power}.


To compare gradient dynamics and EPM, a condition is identified, namely Eq. \eqref{eq.EPM.MO}, which is equivalent to the validity of Maxwell--Onsager reciprocal relations for constitutive relations obtained by EPM. This condition is satisfied for homogeneous entropy productions, but it can be violated in the inhomogeneous case. Validity of Onsager relations is thus guaranteed by EPM near equilibrium, where entropy production is approximately quadratic (i.e. 2-homogeneous), but not generally far from equilibrium.

In summary, the method of EPM has been very successful in engineering and it provides a lot of insight into modeling of complex materials. Reformulating the models as gradient dynamics with dissipation potential leads to a systematic and simple way of obtaining constitutive relations while satisfying additional advantageous properties like Maxwell--Onsager relations. Nevertheless, both methods can positively affect each other.

\section*{Acknowledgement}

We are grateful to O\u{g}ul Esen for discussing the geometric origin of gradient dynamics and to V\'{i}t Pr\r{u}\v{s}a for discussing the method of entropy production maximization.

This work was supported by Czech Science Foundation, project no.  17-15498Y.

Adam Janečka acknowledges the support of Project No. LL1202 in the programme
ERC-CZ funded by the Ministry of Education, Youth and Sports of the Czech Republic.


\appendix

\section{$k$-homogeneous functions}
\label{sec.homogeneous}

A function is $k$-homogeneous if
\begin{equation}
  \label{eq.homogeneous}
  f(\alpha x) = \alpha^k f(x), \qquad \forall \alpha \in \mathbb{R}.
\end{equation}
Taking derivative with respect to $\alpha$ at $\alpha = 1$, we get
\begin{equation}
  x f'(x) = k f(x). 
\end{equation}
Taking derivative of Eq. \eqref{eq.homogeneous} with respect to $x$, we get
\begin{equation}
  \alpha f'(\alpha x) = \alpha^k f'(x),
\end{equation}
thus $f'$ is $(k-1)$-homogeneous. 

When $f$ is a function of several variables, $k$-homogeneous in each variable, the same results hold for partial derivatives and for gradients.

\end{document}